# ROBUST OUTDOOR POSITIONING VIA RAY TRACING

Vladislav Ryzhov

Department of Multimedia Technologies and Telecommunications, Moscow Institute of
Physics and Technology, Moscow, Russia
ryzhov.vp@phystech.edu

## Abstract

*Positioning problem can be solved with several approaches in 5G. This paper introduces RT (Ray Tracing) technique for small-cell Outdoor positioning and tracking for systems with distributed architecture with multipath processing gain. Proposed approach exploits high-resolution angular spectrum estimation and known radio propagation environment. It solves positioning problem robustly even in NLOS cases via CoMP (Cooperative Multi Point) reception and joint processing at cloud CPU. The aim of this paper is to disclose proposed algorithm for NR and compare it to GNSS measurements.*

## 1. INTRODUCTION

Positioning is an old task for wireless communications. The roughest and low-complex methods include identification by Cell-ID and RSS (Received Signal Strength). With high synchronization requirements [1] GPS and TDoA (Time Difference of Arrival) measurements increase positioning accuracy up to dozens of meters. Nowadays, base stations are equipped with large antenna arrays (e.g., 32T32R) and current trend is to split panel into several ones or use single panel with increased number of elements. Under these assumptions angular domain represents the most accurate method for positioning [2]. At the same time the most of classic algorithms suffer from multipath propagation, LOS blockage and variable channel conditions. Another future concept is cooperative MIMO, which is expected to improve spectral efficiency and coverage mostly by increasing number of degrees of freedom and channel diversity [3].

Proposed RT (Ray Tracing) positioning approach exploits multipoint reception and angular spectrum estimation. It is robust to complex propagation conditions and partly benefits from large scattering and multipoint reception. Due to joint RT processing of distributed AoA measurements proposed method is the most suitable for C-RAN (Cloud Radio Access Network) architecture. Also, such approach can be used for real-time CSI (Channel State Information)/ precoder prediction or provide more information for dynamic scheduling for operators.

RT requires high precision maps for accurate localization. In this paper we omit issues related to accumulation data for accurate map reconstruction. With a multiple of automotive sensors, satellite images, cameras, etc. this map can be found. Although our approach does not require additional distributed synchronization, it assumes only intra-panel synchronization and calibration for angular spectrum estimation. Problems related to imperfect calibration, interference directions and inaccurate map are solved via probabilistic weighting and clustering. Anyway, the map inaccuracy is also covered in this paper.

The contribution of this paper is that it proposes a novel approach for precise and robust positioning with a knowledge of propagation conditions and use of distributed RAN architecture. It covers the main aspects of developed framework and, among other things, compares simulations accuracy it with real GNSS measurements.

## 2. SYSTEM MODEL

### 2.1. Channel model

Uplink channel is modelled via RT in certain outdoor environment, obtained from parsed 3D OSM (Open Street Map) with typically located UEs (User Equipment) and RRUs (Remote Radio Unit) from open-source maps with mobile operator towers and their frequency bands.

### 2.2. Layout

Cooperating RRUs (with $120^0$ sector antenna) are placed to satisfy ISD $\approx$ 150 m, tilt $\approx 10^0$, height $\approx$ 15-25 m. UEs are evenly spaced in the street at height of 1.5 m. For simplicity each UE is equipped with one antenna with uniform pattern, while RRU has (4x8) horizontal antenna array with 3GPP antenna element radiation pattern [4].

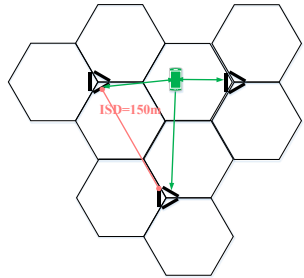

**Figure 1.** *Layout with 3-sector antenna.*

### 2.3. Signals

Each UE is configured [5] to send SRS (Sounding Reference Signal) in TDD mode with granularity K_TC = 2 and N_RB = 272 PRBs.

### 2.4. Overall positioning model

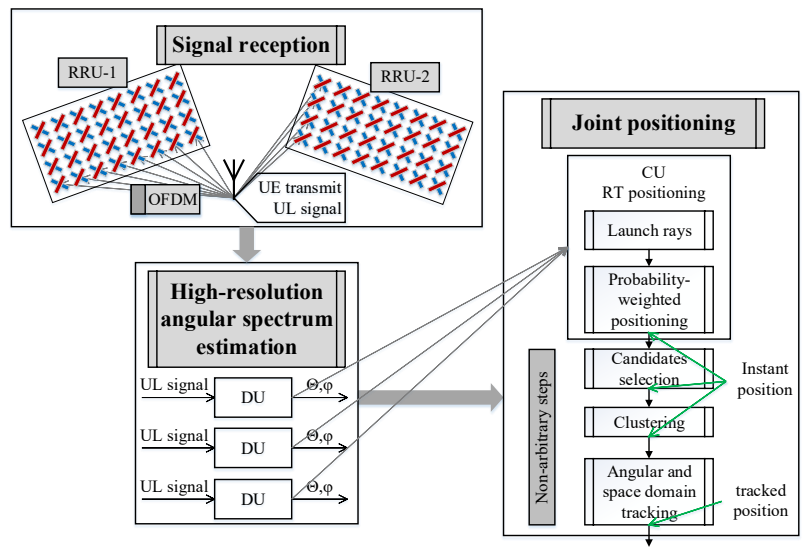

**Figure 2.** *System model for positioning.*

Multiple distributed RRUs cooperatively receive SRS, passed through RT channel. After that each DU (Distributed Unit) independently finds AoAs (Angles of Arrival) from estimated angular spectrum. These estimates are reported to CU (Central Unit), which contains precise outdoor model. CU localize UE from current instantaneous measurement via RT search in a direction of received angles. At this step position represents probabilistic estimate. Discarding and clustering of estimated positions is done to cope with outliers from imperfect angle estimation and map errors. Also, CU track measurements in angular domain and expectation of positions for each cluster in space domain in order to additionally discard false positions and smooth UE trajectory. To keep trade-off between required accuracy and complexity we are able to exchange additional clustering or tracking complexity for higher accuracy.

## 3. RT Channel Model

RT channel modelling has a good accuracy [7] and papers [8-10] describes RT positioning. Developed C++ RT simulator has two main operating options: channel modelling and positioning. In both cases it contains the following propagation patterns:

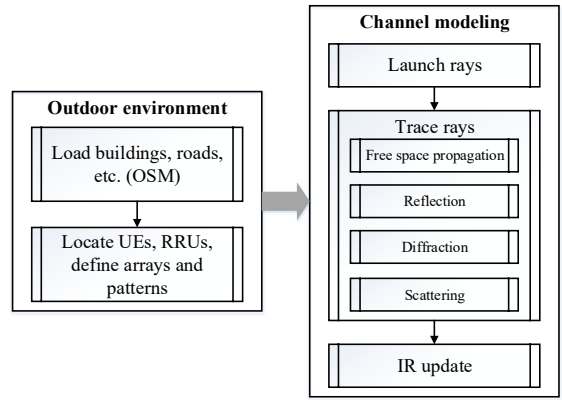

**Figure 3.** *Concept of developed channel modelling via RT.*

- Free space propagation

Solution of Helmholtz wave equation for point source in far field is a spherical wave, which defines phase and amplitude of propagating ray:

$$E(r, \theta, \varphi) = \frac{E_0(\theta, \varphi)e^{-ikr}}{r} \tag{1}$$

- Reflection

When interfered with by building, ground, etc. ray is reflected according to geometrical optics and reflection coefficients. From a Fresnel's equation reflection coefficients can be found from angles of incidence and reflection and electrical properties of material:

$$R_\perp = \left| \frac{n_2 \cos(\theta_i) - n_1 \cos(\theta_t)}{n_2 \cos(\theta_i) + n_1 \cos(\theta_t)} \right|^2$$

$$R_{||} = \left| \frac{n_2 \cos(\theta_t) - n_1 \cos(\theta_i)}{n_2 \cos(\theta_t) + n_1 \cos(\theta_i)} \right|^2 \tag{2}$$

- Scattering

Due to the roughness of surface, ray is scattered around the direction of reflection. Authors of [6] performed measurement campaign for scattering pattern of typical buildings and concluded that single-lobe directive model was the closest:

$$|\overline{\mathbf{E_S}}|^2 = \mathbf{E_{S0}^2} \left( \frac{\mathbf{1 + \cos \psi_R}}{\mathbf{2}} \right)^{\mathbf{\alpha_R}}, \mathbf{\alpha_R = 4} \qquad (3)$$

- Diffraction

Paper [11] defines the most common Knife Edge Diffraction (KED) model:

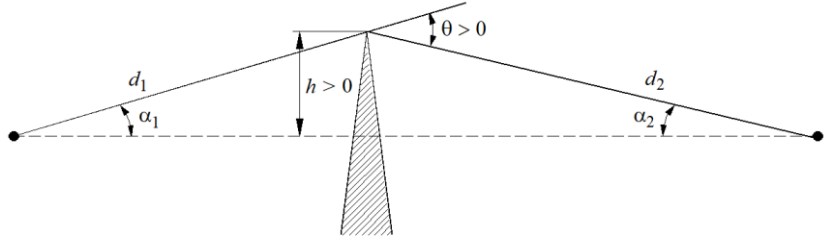

**Figure 4.** *For definition v [11].*

$$\mathbf{v = h} \sqrt{\frac{\mathbf{2}}{\mathbf{\lambda}} \left( \frac{\mathbf{1}}{\mathbf{d_1}} + \frac{\mathbf{1}}{\mathbf{d_2}} \right)} \qquad (4)$$

where electric field can be found from:

$$\mathbf{F(v) = \frac{E_{diffracted}}{E_0} = \frac{1 + i}{2} \int_v^\infty e^{-\frac{i \pi t^2}{2}} dt} \qquad (5)$$

and approximation ($\mathbf{v > -0.7}$) for diffraction loss in dB:

$$\boldsymbol{J(v) = 6.9 + 20 \log \left( \sqrt{(v - 0.1)^2 + 1} + v - 0.1 \right)} \qquad (6)$$

For detailed maps prediction accuracy of RT channel modelling is high [12]. The main hardship is complexity, which can be reduced with minimum storage and computations techniques. For these reasons the following methods were used for acceleration of tracing in C++ simulator:

- BVH (Bounding Volume Hierarchy) is a method [13] used for compact storage and reduced intersection testing. In order to decrease number of objects to compare, each object and UE are mapped to a spatial grid, so that ray is interfered only with objects in current grid cell.

- Parallelepiped wrappers are utilized to perform fast initial intersection test and discard objects without complex surface processing.

- Fast implementation of 3D geometry for ray tracing [14-15] with an extensive application of fast ray-triangle intersection [16].

## 4. POSITIONING

### 4.1. Independent AoA estimation

Firstly, each RRU estimate independently high-resolution angular spectrum from received SRS, e.g., via MUSIC algorithm [17].

$$\mathbf{S(\varphi, \theta) = \frac{1}{a^H(\varphi, \theta) VV^H a(\varphi, \theta)}} \qquad (7)$$

$$a(\boldsymbol{\varphi}, \boldsymbol{\theta}) = \Delta_x^i \cos(\boldsymbol{\theta}) \sin(\boldsymbol{\varphi}) + \Delta_v^i \cos(\boldsymbol{\theta})\cos(\boldsymbol{\varphi}) + \Delta_z^i \sin(\boldsymbol{\theta}) \qquad (8)$$

$$\begin{pmatrix} \Delta_x^i \\ \Delta_y^i \\ \Delta_z^i \end{pmatrix} = \begin{pmatrix} (i_h - 1)d_h\cos(\boldsymbol{\varphi_h}) + (i_v - 1)d_v\sin(\boldsymbol{\varphi_h})\sin(\boldsymbol{\theta_v}) \\ (i_h - 1)d_h\sin(\boldsymbol{\varphi_h}) - (i_v - 1)d_v\cos(\boldsymbol{\varphi_h})\sin(\boldsymbol{\theta_v}) \\ (i_v - 1)d_v\cos(\boldsymbol{\theta_v}) \end{pmatrix} \qquad (9)$$

where $V$ – matrix whose columns are noise eigenvectors, $a(\varphi, \theta)$ – steering vector of antenna panel with $\boldsymbol{\theta_v}$ – tilt and $\boldsymbol{\varphi_h}$ – rotation of antenna array in horizontal plane. AoAs are estimated from spectrum $\mathbf{S}(\boldsymbol{\varphi}, \boldsymbol{\theta})$ via CA-CFAR (Cell Averaging Constant False Alarm Rate) detector:

$$\mathbf{I}[[\boldsymbol{\varphi_n}, \boldsymbol{\theta_n}]] = \mathbf{I}\left[ P(\boldsymbol{\varphi_n}, \boldsymbol{\theta_n}) > \frac{C_{threshold}}{N_\varphi N_\theta} \sum_{\substack{i=-\frac{N_\varphi}{2} \\ i \neq 0}}^{\frac{N_\varphi}{2}} \sum_{\substack{k=-\frac{N_\theta}{2} \\ k \neq 0}}^{\frac{N_\theta}{2}} P(\boldsymbol{\varphi_{n-i}}, \boldsymbol{\theta_{n-k}}) \right] \qquad (10)$$

## 4.2. Joint estimation of spatial probability density

Cloud CU receives noisy estimates of angles $\boldsymbol{\Omega^{est}} = \{\boldsymbol{\varphi_n}, \boldsymbol{\theta_n}\}$ from each DU and search for positions by launching rays in close directions $\boldsymbol{\Omega^{est}} + \boldsymbol{\Omega^{bias}}$. Intersections of multiple rays from several RRUs define possible emitter locations. Each of positions $\{\mathbf{P_{x,y,z}^i}\}_{i=1}^{N_{pos}}$ is associated with probabilities of received rays and the expectation maximization leads to:

$$\mathbf{P_{x,y,z}} = \frac{1}{\sum_i \mathbf{w_i}} \sum_i \mathbf{w_i} \, \mathbf{P_{x,y,z}^i} \qquad (11)$$

Bayesian approach can be used to cope with angle estimation errors caused by interference and residual inter-RRU calibration errors:

$$\mathbf{P(\Omega_{bias}|\Omega_{est})} = \frac{\mathbf{P(\Omega_{est}|\Omega_{biased})P(\Omega_{biased})}}{\mathbf{P(\Omega_{est})}} \qquad (12)$$

where $\boldsymbol{\Omega_{est}}$ – biased estimation of angle (azimuth or elevation) received from DU, $\boldsymbol{\Omega_{bias}}$ – small angle deviation from received direction. $\mathbf{P(\Omega_{biased})}$ – unique for certain RRU antenna panel and does not depend on UE (e.g., $\mathbf{P(\Omega_{biased})} \sim \mathbf{N}(0, \frac{|\Omega_{biased}^{max}|}{3})$)), $\mathbf{P(\Omega_{est}|\Omega_{biased})}$ – related to estimation from angular spectrum (peak value and peak width) for each found ray.

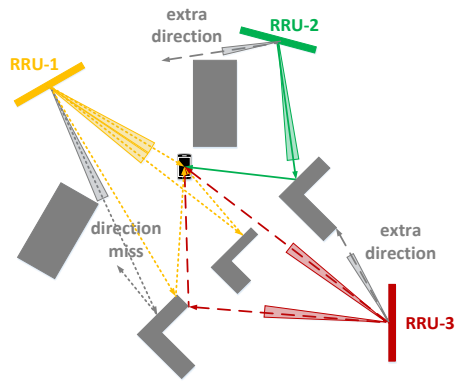

**Figure 5.** *Ray search at CU via RT and known propagation environment.*

$$\mathbf{w_i} = \sum_{r=1}^{N_{rays}} \mathbf{P}(\Omega_{est}|\Omega_{biased})\mathbf{P}(\Omega_{biased}) \qquad (13)$$

At the end of this step each found position $\mathbf{P_{x,y,z}^i}$ is described with its probability $\mathbf{w_i}$. From formula (11) we may find single state maximum expected position. For some cases (e.g., UE has connection to several RRUs) this estimate will be enough to produce the only dense cluster with weighted centre very close to real position.

## 4.3. Optimal position selection

At this stage development of algorithms for joint processing of positions and RT model outputs has a high prospects and research possibilities. RT outputs may include number of rays, number of reflections per each ray, closeness ray to reflecting/diffracting edges, LOS/NLOS recognition, number of BSs, power level indicators, specific hardware limits of antennas and others. Processing of parameters above requires much fewer computing resources than application of RT model. At the same time such algorithm can enhance selection of the most probabilistic positions via recognition of typical patterns. Performance gain can be achieved by taking into account correlations between algorithm positioning errors and physical propagation conditions which are in general related to side RT outputs. As you can see in section 5, even the simplest parameters such as number of BSs or LOS/NLOS indicator are good predictors of errors.

## 4.4. Clustering of probabilistic estimates of position

In general case erroneous angle acquisition may cause tracking extra directions (e.g., in music algorithm it is caused by interference) or direction miss (e.g., small search area). These errors lead to decomposition of all positions into several remote clusters. Despite probability of fake clusters is typically low, it slightly decreases accuracy of estimation. In order to increase accuracy, we need to perform simple clustering of all positions and exclude fake clusters from consideration.

| **Clustering algorithm** |
| --- |
| 1. Estimate table of distances. Initialize minimum distance between clusters $d_{cluster}$. <br> 2. Perform DF (Depth-First search) of connected subgraphs with condition of connectivity: $d_{i,k} < d_{cluster}$ <br> 3. Find clusters' centers from formula (11) and corresponding variances. |

First step of proposed clustering approach has complexity $\mathbf{O(N_{pos}^2)}$, while the second and third steps are linear. In order to exclude storage of large table (memory $\sim \mathbf{O(N_{pos}^2)}$) of distances this algorithm can be replaced with similar with storage of logical vector of added positions (memory $\sim \mathbf{O(N_{pos})}$).

If the following tracking is not being exploited the most likely cluster is used for selection of optimal position (weighted with probabilities / max posterior position / combining / more sophisticated with RT outputs).

## 4.5. Tracking in Angular Domain

AoA knowledge is equivalent to knowledge about position for accurate 3D maps. Tracking of direct measurement (angles) is more appropriate than mediated because the error distribution for the first one is more understandable, while the second one may produce biased trajectory. Due to calibration residuals, interference and scattering estimate for azimuth/elevation is close to mixture of uniform and normal/cauchy distributions or contaminated ones. For these reasons L2 optimization is less appropriate for tracking [18] rather then Lp (1<p<2) which is closer to median (L1). Another reason to choose L1 were simulations, resulted in better error reduction mainly due to ability to cope with outliers.

| **AoA tracking algorithm** |
| --- |
| <u>Do for each TTI with new AoA estimate:</u><br>1. Start trend from new angle estimate.<br>2. Check whether new estimate can be used to update current trends.<br>3. Fit each single trend to a new estimate.<br>4. Check for redundant trends.<br>5. Erase noisy, redundant and outdated trends, shrink buffers.<br>6. Predict angles with corresponding trend deviations. |

$$\cos(\Omega_{TTI_i} + \Delta\Omega) \approx \cos(\Omega_{TTI_i}) - \sin(\Omega_{TTI_i})\Delta\Omega \qquad (14)$$

Figure 6 represents assumptions for developed AoA tracking system. It includes linear change of noised AoA between TTI (from steering vectors (9) and long distance to tracking object (14)), AoA estimation miss, false AoA detection, sudden AoA change (UE changed travelling direction or got around the corner of the building) and irregular TTI structure. AoA tracking is done for each antenna array and azimuth/elevation independently. Lp fit for single trend is described in [19].

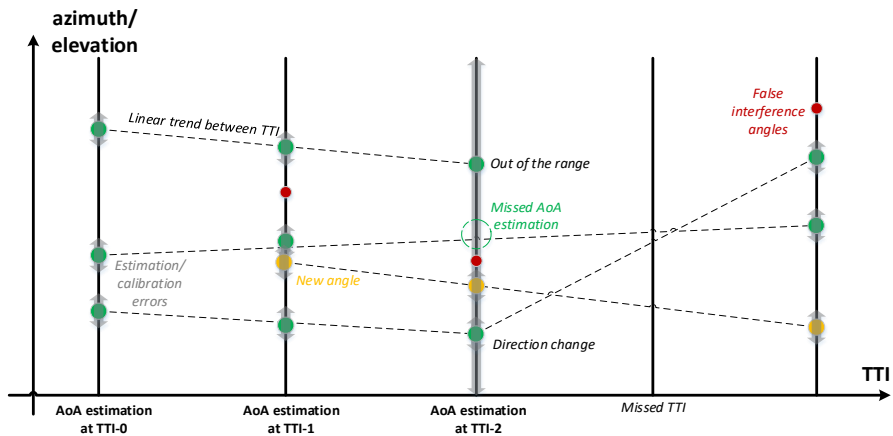

**Figure 6.** *Assumptions for angle tracking model.*

Angle predictions are used then for RT positioning with probabilistic weights (13) and additional weights from trend deviations from azimuth and elevation:

$$w_{ray} = \frac{N_{buffer}}{\sum_{i=1}^{N_{buffer}} |\Omega_i - a_{trend} \cdot TTI_i - b_{trend}|^p} \qquad (15)$$

The same tracking system without separation for several trends is used for main cluster tracking. Figure 7 reveals the idea for cluster tracking.

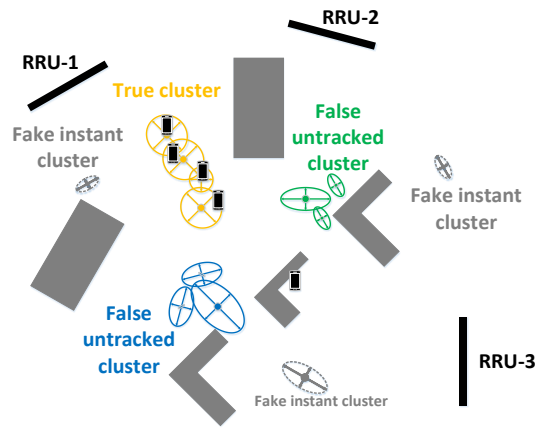

**Figure 7.** *Representation of clusters tracking.*

## 4.6. Accuracy control

Estimation both position and predicted error is important for combining RT with other positioning approaches. For RT great errors in instant position estimation can be controlled by comparison with measurements for received power and time differences (not taken into account in this paper). Another way is to exploit RT outputs as described in section 4.3. E.g., miss can be estimated via variance of maximum likelihood points:

$$I\{P_{x,y,z} = miss\} = I\{D[P_{x,y,z}^i] > \varepsilon_0\} \qquad (16)$$

## 5. SIMULATION RESULTS

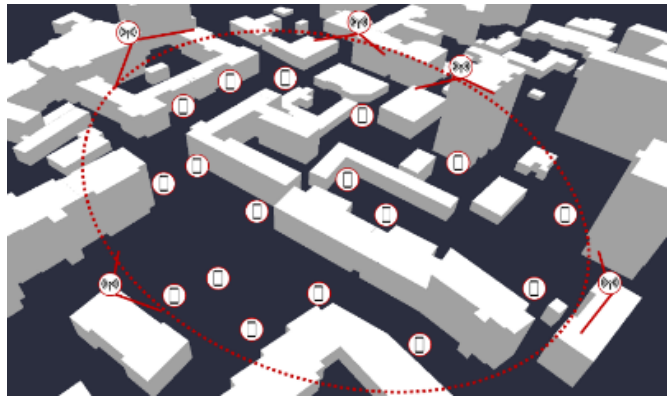

**Figure 8.** *Example of studied Outdoor Environment.*

## 5.1 Positioning accuracy

Figure 8 represents simulated Urban Environment, parsed from OSM map. Base stations' antenna arrays were placed at the top of the buildings. Figure 9 shows how RT localization principle works:

- base station estimates angular spectrum and detects 3 directions;
- CRAN performs RT approach and detects position of UE

In case of accurate localization estimated CIR might be improved. Very close taps (e.g., LOS and ground reflection) might be distinguished. This result might be also used for more accurate beamforming calculation for open-loop systems.

Figure 10 shows CDF of positioning error of single-state position estimation. It is noticeable that impairments that lead to erroneous angle acquisition also cause dramatic decrease of positioning accuracy. Second graph demonstrates that low accuracy of phase calibration of antenna panel for NLOS cases makes RT positioning without clustering and position selection not robust.

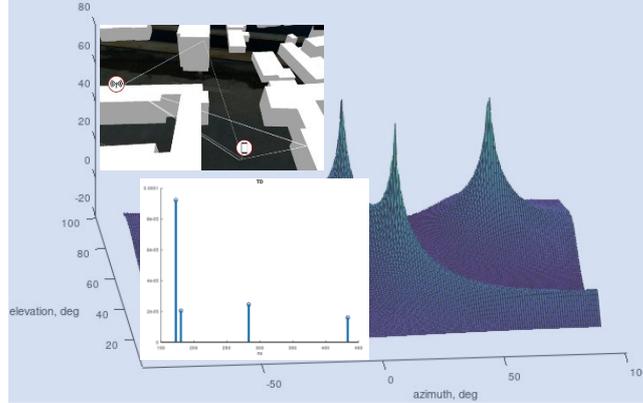

**Figure 9.** *Example of high precision angular spectrum estimation and ray tracing for positioning and CIR reconstruction.*

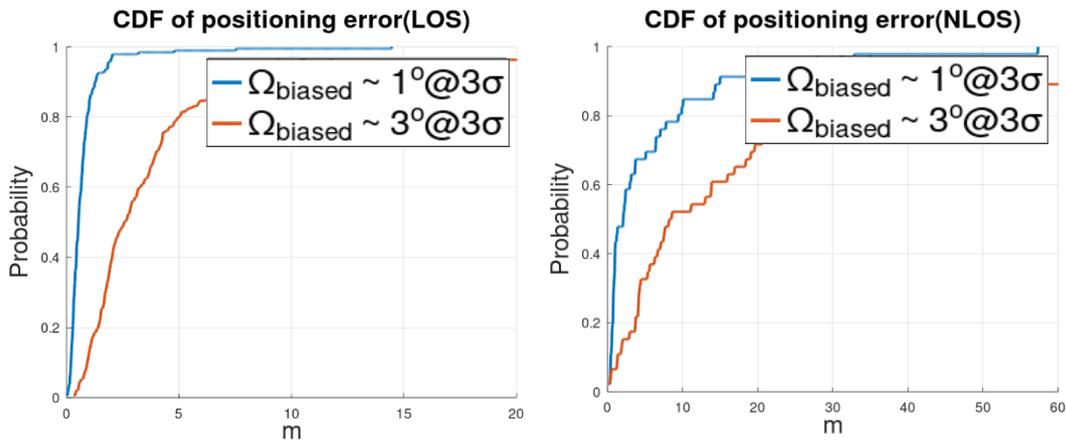

**Figure 10.** *CDF of error of joint estimation of the most likely position (after step 4.2).*

Table 1 demonstrates dependence of positioning error on the number of base stations used for SRS acquisition. The high positioning error occur primarily due to the points of poor radio coverage, where the use of several base stations is limited. If we take into account that such points are known in advance, then it is possible to control this error. Research has shown that such areas are related to reception of signal from single collocated antenna array with very low SNR and sparse NLOS channel.

**Table 1.** *Localization accuracy with dedicated maximum number of base stations.*

| LOS&NLOS +/- $1^0@3\sigma$, without tracking | | | | | |
|---|---|---|---|---|---|
| Nbs | 1 | 2 | 3 | 4 | 5 |
| 90% percentile for absolute error | 15m | 2m | 1.6m | 1.3m | 1m |

Figure 11 represents CDF of positioning error (+/- $2^0$) with clustering and rejection of position candidates by the only flexible threshold for number of rays. Previously it was mentioned that impairments can cause fake remote position candidates to appear and decrease accuracy significantly. Such misdetection was solved with selection of the most likely cluster (e.g., in figure 12). Inside cluster several approaches can be used:

- Position with maximum posterior probability
- Most expected position, averaged in cluster

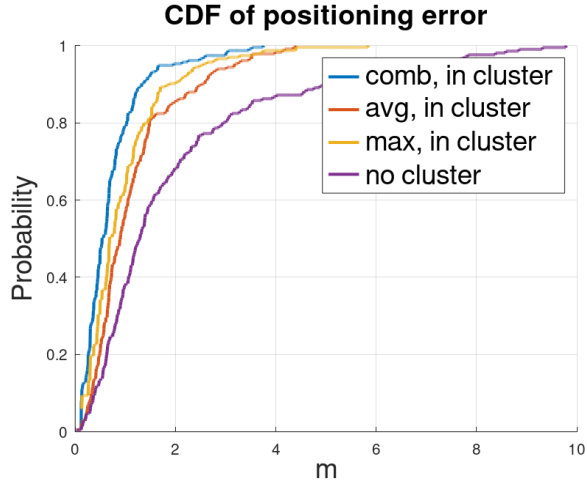

**Figure 11.** *CDF of error of joint estimation of expected position with main cluster selection (after step 4.4).*

Both of these strategies are presented in graph as well as their combination. Optimal clustering provides positioning error with Pr = 0.9 less than 1.3m for typical mixed LOS/NLOS Urban Outdoor scenario while without clustering 1.3m is reached with Pr = 0.5. In case of more accurate phase synchronization (+/- $1^0@3\sigma$) at the same probability level localization accuracy reaches 0.9m, which perfectly fits 5G NR requirements for accuracy [20-21].

True positioning error is highly correlated to interim assessment of position variance among a set of possible positions with Pearson correlation coefficient = 0.7, which means that the most of predicted positions can be estimated with very close errors. Better error control as mentioned earlier can be achieved by matching positions with exact patterns from RT outputs.

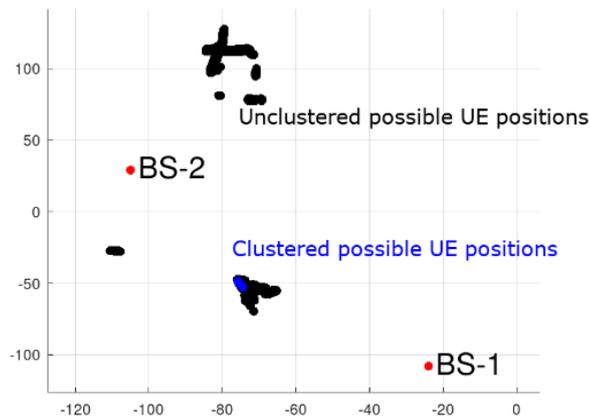

**Figure 12.** *Example of clustering of candidates for UE position*

Figures 13-14 represent example of UE trajectory in NLOS and corresponding tracking error. Highest errors (above 1m) happen during direction change in sparse NLOS conditions. After UE enters LOS area (after 310 TTI, TTI = 10 frames) tracking adapts much faster and error mostly does not exceed 1m after convergence.

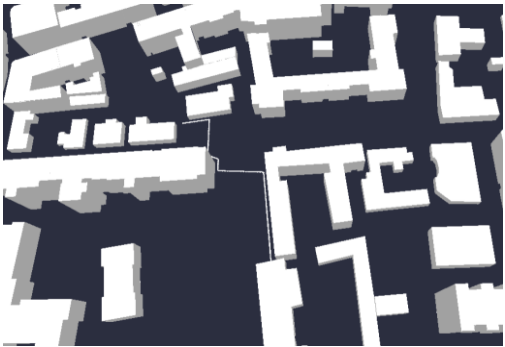

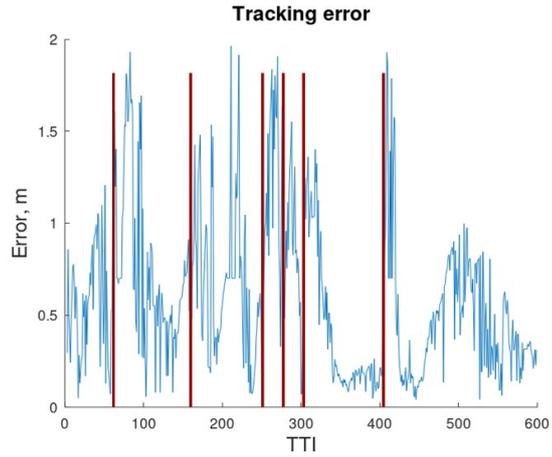

**Figure 13.** *Example of UE trajectory.*

**Figure 14.** *Absolute tracking error across the track (Figure 13). Vertical lines correspond to the moments when UE turns in another direction.*

Figure 15 shows UE trajectory in azimuth azimuth/time plane from single base station. Predicted (green) have low level of outliers. Such approach can also be useful for management of frequent UE tracks to enhance positioning via storage of previous measurements and use them to predict optimal multiuser beamforming.

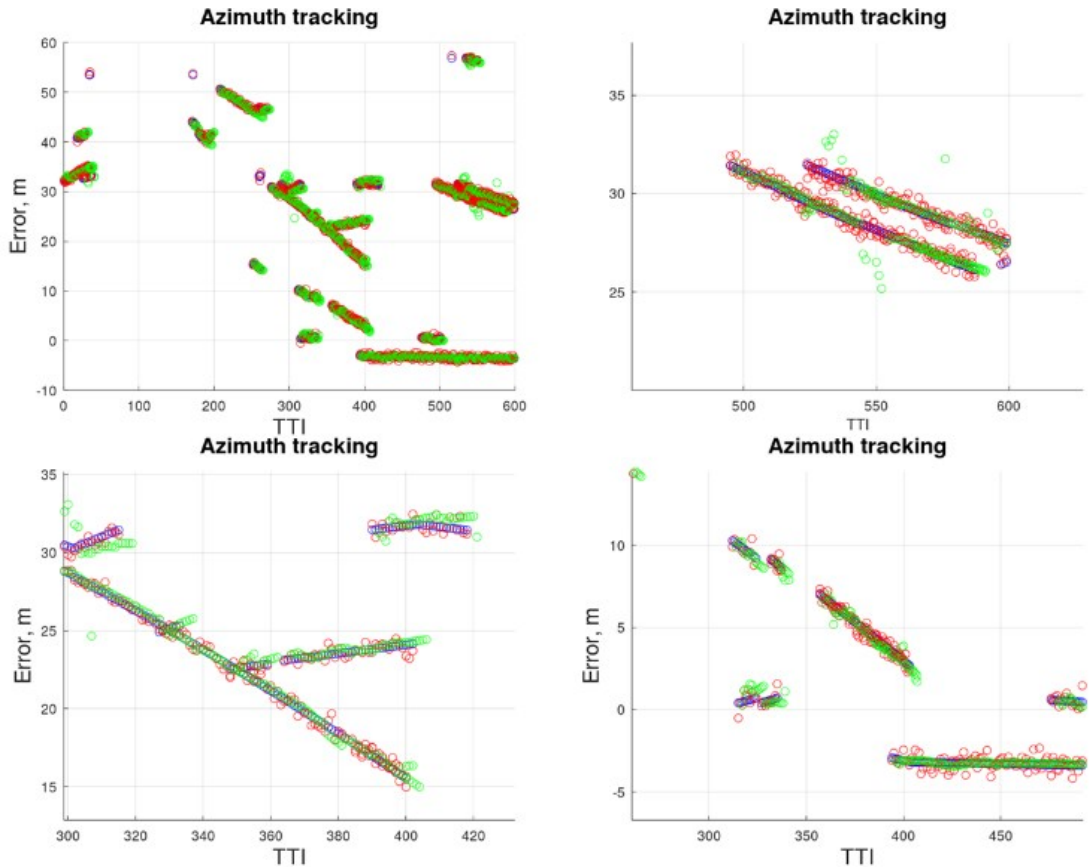

**Figure 15.** *Azimuth tracking of UE trajectory. Blue – true AoA, red – measurement (interference not shown), green – predicted.*

## 5.2. Impact of Inaccurate Maps

Initially the most concerns about positioning accuracy were related to the accuracy of the digital map. These were resolved with RT positioning in the map, where properties of materials (reflection coefficients, scattering model), buildings sizes and positions are pre-set with errors. Results are summarized in table 2. These results are very close to estimate by Monte Carlo method if simulated with known average number of reflections and rays to each UE in 2D space.

**Table 2.** *Contribution of map inaccuracy to localization error.*

|  | Deviation from true value | median | 90% percentile |
|---|---|---|---|
|  | no | 0.7 m | 1.3 m |
| Transmission coefficient | $Uniform(50\%, 200\%)$ | 1 m | 1.7 m |
|  | $Uniform(80\%, 150\%)$ | 0.8 m | 1.5 m |
| Building, walls position | $N(0\ m, 1\ m)$ | 1.2 m | 2.1 m |
|  | $N(0\ m, 2\ m)$ | 1.5 m | 2.5 m |

The most of increase of position error from inaccurate map came from UE with a little number of rays or served by the only base station. Also ray search area has to be increased to partly mitigate error.

## 5.3. Comparison of Measurements and Simulations

Simulation results above were compared with real GPS measurements in the same outdoor positions as simulated. About 1000 measurements were done both via Samsung Galaxy Note 8 (*GPS Test* software) and iPhone SE (*GPS Data Smart* software). The following metrics were collected for each measurement:

- Horizontal accuracy
  Internal software estimate of positioning accuracy corresponding to 68% confidence interval.
- Measured accuracy
  Absolute difference of positions in Google or Yandex maps and real environment (compared to surrounding buildings)

Table 3 provides results of experiment and compares with simulations (with combined cluster selection, but without tracking). On average proposed approach has a several times higher accuracy than GNSS one.

**Table 3.** *Comparison between accuracy of measurements and simulations.*

|  |  | median | 90% percentile |
|---|---|---|---|
| Note 8 | Horizontal accuracy | 5.2 m | 8.9 m |
|  | Measured accuracy | 4.7 m | 8.6 m |
| iPhone SE | Horizontal accuracy | 5.0 m | 7.9 m |
|  | Measured accuracy | 7.6 m | 11..2 m |
| Simulated | $1^0@3\sigma$ | 0.5 m | 1.2 m |
|  | $3^0@3\sigma$ | 0.7 m | 1.5 m |

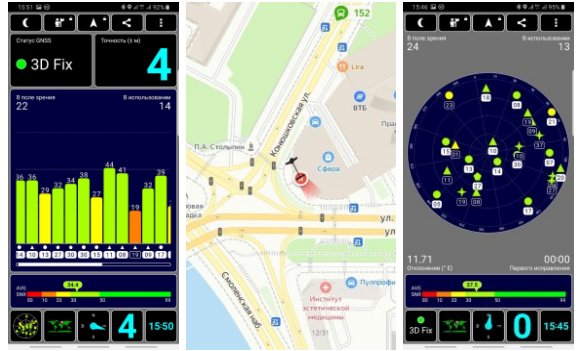

**Figure 16.** *Example of usage GPS Test and Yandex Maps for single measurement.*

## 6. CONCLUSIONS

This paper covered RT positioning technique for Urban Outdoor scenario and supplemented previously studied positioning algorithms. It was shown that impairments and interference cause pure RT positioning to fail in NLOS. Developed clustering approach for probabilistic candidates for positions mostly mitigated this error and provided almost the same accuracy as for LOS case. Proposed positioning framework was compared to measurements and demonstrated much better accuracy. For 90% Outdoor cases RT positioning may provide ~0.9-2.5m accuracy, which is enough for the most 5G use cases. Also, it was shown that that proposed algorithm is robust and able to work even in low precision maps.

Developed low complex tracking system in angular domain demonstrated ability to efficiently cope with angle acquisition outliers. This can be used to decrease allocated resources per UE or improve scheduling.

RT positioning has a good internal assessment of error, which enables to combine it with other positioning systems.

## FUTURE RESEARCH

In the future it is expected to develop prediction model for probabilistic SU/MU-MIMO hybrid precoder with use of RT and angle tracking. Preliminary simulations have shown that such approach is possible to save up to dozens of percent of sum rate for moving pedestrian in urban scenario.

## DISCUSSION ABOUT IMPLEMENTATION AND PROBLEMS

Proposed RT approach for positioning is beneficial for future RAN at least because it produces new independent estimates for joint filter exploiting GPS, sensors, TDoA from base stations, etc. It does not burden edges with additional large computational resources of traffic and transfer the main load to server.

As mentioned above, RT positioning in general case requires detailed maps. These may be used from external reliable sources. Anyway, below are some ideas for inaccurate map calibration:

- RT approach can be used to shift biased buildings in digital maps. Some areas of maps are covered by several base stations. For multiple UE positions corresponding to the same reflectors for several distributed antenna panels multiple biases in high-resolution FIR taps can be used. Synchronization errors are decreased by multiple pairs of reflectors for different UE positions. Shifting of object borders seems to be solved with GAN models.

- Another approach is offline and exploit RTK receivers for digital map calibration. In this case sets of low cost UEs are replaced with single accurate reducing computation complexity.

Another important issue to mention is fast, low-complex and memory-efficient RT model. For the only purpose of positioning, it can be effectively stored in database containing sets <angle, position, RT outputs> where RT outputs links to another database <index, object description>. RT outputs are any important parameters that can enhance optimal position selection stage or be used for other prediction needs. Many simulated cases have shown that via C++ from several seconds to several days (depends on accuracy, details, map size) may be required to simulate with a single thread. These seems to be opportunistic, because it should be done once to fill the database of several Gb. Also due to rays are independently propagate these computations can be easily parallelized and also some parts of database could be upgraded if propagation environment has changed without complete computation.

## ACKNOWLEDGEMENTS

This paper will not be possible without support of my supervisor Anton Laktyushkin. His knowledge and critics let me avoid many common mistakes in this work. I would also like to thank all the colleagues from Huawei Moscow Research Center with whom I had the opportunity to work.

## AUTHORS

**Vladislav Ryzhov** is a researcher and developer in the field of LTE/5G. He received B.S. degree in MIPT in the field of applied mathematics and physics in 2021 and continued his studies towards M.S. He has an industry experience in Huawei Technologies Co. Ltd. and Yadro.

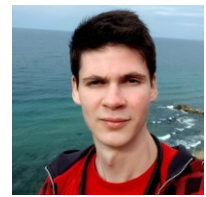